\documentclass[fleqn,10pt]{wlscirep}
\usepackage{bm}

\graphicspath{{../Figures/}}

\usepackage{epstopdf}
\usepackage{epsfig}
\usepackage{pdfpages}

\title{Control of complex networks requires both structure and dynamics}

\author[1,2+]{Alexander J. Gates}
\author[1,2,3*]{Luis M. Rocha}
\affil[1]{School of Informatics and Computing, Indiana University, Bloomington, IN, USA}
\affil[2]{Program in Cognitive Science, Indiana University, Bloomington, IN, USA}
\affil[3]{Instituto Gulbenkian de Ciencia, Oeiras, Portugal}
\affil[+]{ajgates@indiana.edu}
\affil[*]{rocha@indiana.edu}


\begin{abstract}
The study of network structure has uncovered signatures of the organization of complex systems. However, there is also a need to understand how to control them; for example, identifying strategies to revert a diseased cell to a healthy state, or a mature cell to a pluripotent state.
Two recent methodologies suggest that the controllability of complex systems can be predicted solely from the graph of interactions between variables, without considering their dynamics: structural controllability and minimum dominating sets.
We demonstrate that such structure-only methods fail to characterize controllability when dynamics are introduced. 
We study Boolean network ensembles of network motifs as well as three models of biochemical regulation: the segment polarity network in \textit{Drosophila melanogaster}, the cell cycle of budding yeast \textit{Saccharomyces cerevisiae}, and the floral organ arrangement in \textit{Arabidopsis thaliana}.
We demonstrate that structure-only methods both undershoot and overshoot the number and which sets of critical variables best control the dynamics of these models, highlighting the importance of the actual system dynamics in determining control.
Our analysis further shows that the logic of automata transition functions, namely how canalizing they are, plays an important role in the extent to which structure predicts dynamics.

\end{abstract}
\begin{document}

\flushbottom
\maketitle

\thispagestyle{empty}

% currently 4613 words (they claim the max is now 4500, not the page limits)
\section*{Introduction}
Complex systems are typically understood as large nonlinear systems.
Their organization and behavior can be modeled by representations such as graphs and collections of automata.
Graphs are useful to capture the \emph{structure} of interactions between variables: the static organization of complex systems.
However, nodes representing variables in graphs lack intrinsic dynamics.
The simplest way to study nonlinear \emph{dynamics} is to allow network nodes to have discrete states and update them with automata; for instance, Boolean Networks (BNs) are canonical models of complex systems which exhibit a wide range of interesting behaviors \cite{Kauffman1993}.

The study of network structure has uncovered several organizing principles of complex systems --- such as scale-free networks and community structure --- and how they constrain system behavior, without explicit dynamical rules for node variables \cite{Newman2003}.
There is, however, a need to \emph{control} complex systems, in addition to characterizing their organization.
This is particularly true in systems biology and medicine, where increasingly accurate models of biochemical regulation have been produced. \cite{Huang1999,Barabasi2004,Zhu2007,Assmann2009}
More than understanding the organization of biochemical regulation, we need to derive control strategies that allow us, for instance, to revert a mutant cell to a wild-type state \cite{MarquesPita2013}, or a mature cell to a pluripotent state \cite{Wang2011}.
While the identification of such control strategies occurs for a given model, not the real system, predictions from control theory can be used for model verification and thus also aid the separate question of the accuracy of that model in predicting the real system.

Network structure has been reported to predict properties of dynamics, such as the synchronization of connected limit-cycle oscillators \cite{Strogatz2001}, or the likelihood of robust attractors \cite{Klemm2005}.
On the other hand, there are important system attributes which depend on dynamical characteristics of variables and their interactions; e.g. the critical transition between ordered and chaotic dynamics in BNs depends both on structural (mean connectivity) and dynamical properties of nodes (bias and canalization) \cite{Shmulevich2005,Nykter2008,Hossein2014,MarquesPita2015}.
Indeed, we already know that such dynamical properties strongly impact the stability, robustness, and controllability of existing models of gene regulation and biochemical signaling in a number of organisms \cite{MarquesPita2013,Shmulevich2003,Kauffman2004,Li2004,Gershenson2006}.
Therefore, a question of central importance remains: \emph{How well does network structure predict the dynamics of the underlying complex system, especially from the viewpoint of control?}

Recently, two related methodologies were used to predict the controllability of complex networks based solely on network structure without consideration of the dynamical properties of variables:
\emph{structural controllability} (SC) \cite{Lin1974,Liu2011} and \emph{minimum dominating set} (MDS) \cite{Nacher2012,Nacher2013}.
Both techniques reduce dynamical systems to graphs where edges denote an interaction between a pair of variables.
Using only graph connectivity, the goal is to identify a minimal set of \emph{driver variables} (a.k.a.\ driver nodes) which can fully control system dynamics \cite{Valente2012}.

SC assumes that, in the absence of cycles, a variable can control at most one of its neighbors in the structural interaction graph \cite{Lin1974,Liu2011}.
The influence from an intervention on a node then propagates along a backbone of directed paths, where the number of necessary paths to cover the network dictates the minimum set of driver variables (see Supplemental Material, SM).
Cycles are considered to be self-regulatory and do not require an external control signal.
SC has become an influential method, having been used to suggest that biological systems are harder to control and have appreciably different control profiles than social or technological systems \cite{Egerstedt2011,Ruths2014}.
The methodology has also been used to identify key banks in interbank lending networks \cite{Delpini2013}, and to relate circular network motifs to control in transcription regulatory networks \cite{Osterlund2015}.
However, despite its successful characterization of observability (a dual notion to controllability) in several nonlinear dynamical systems \cite{Liu2013},
SC's application to models of biological and social systems has been heavily critiqued due to its stringent assumptions \cite{Muller2011,Cowan2012,Sun2013}.

MDS starts from the different assumption that each node can influence all of its neighbors simultaneously, but this signal cannot propagate any further.  Driver variables are then identified by the minimal set such that every variable is separated by at most one interaction \cite{Nacher2012, Nacher2013}.
It has been used to identify control variables in protein interaction networks \cite{Wuchty2014} and characterize how disease genes perturb the Human regulatory network \cite{Wang2015}.

Because both MDS and SC use only the interaction graph of complex systems, unless otherwise specified, we use \emph{structural control} to refer to both methods.
Since these methods are increasingly used in a variety of scientific domains, it is important to study how much network structure predicts the controllability of realistic, nonlinear dynamical systems.

Here, we explore this problem using ensembles of BNs.
These canonical models of complex systems are defined by a network of interconnected automata (the structure), and exhibit a wide range of dynamical behaviors \cite{Kauffman1993}.
They have been used to model biochemical regulation in organisms, where dynamical attractors represent cell types, disease and healthy states \cite{Assmann2009, Bornholdt2008}.
It is well known that when the set of system variables is large, enumeration of the state-spaces of BNs becomes difficult, making the control problem for general deterministic BNs computationally intractable (NP-hard) \cite{Akutsu2007}.
However, for small systems we can fully enumerate the state-space and compute the actual controllability (as measured by three proposed measures of controllability) for parameterized ensembles of BNs.

Our analysis is not meant to introduce alternative techniques to uncover control variables in BNs, since methods based on system dynamics already exist \cite{MarquesPita2013,Wang2011,Langmead2009,Cheng2009,Srihari2013,Jia2013, Li2015,Lu2015,Motter2015}.
The goal is to quantify the discrepancy between control as uncovered by approximate methods that use structure alone, from how actual control unfolds in BNs.
Additionally, we characterize critical variables for the control of three models of biochemical regulation: the single-cell segment polarity network in \emph{Drosophila melanogaster}, the eukaryotic cell cycle of budding yeast \emph{Saccharomyces cerevisiae}, and the floral organ arrangement in the flowering plant \emph{Arabidopsis thaliana}.
Our results demonstrate that network structure is not sufficient to characterize the controllability of complex systems; predictions based on structural control can both under- and over-estimate the number and set of necessary driver variables.
Therefore, previous assertions about the controllability of biochemical systems reached from analyses based on structural control methods do not offer a realistic portrayal of control \cite{Egerstedt2011,Ruths2014}.

%%%%%%%%%%%%%%%%%%%%%%%%%%%
\section*{Quantifying Control in Boolean Networks}
%%%%%%%%%%%%%%%%%%%%%%%%%%%

%%%%%%%%%%%%%%%%%%%%%%%%%%%
\subsection*{Background}
%%%%%%%%%%%%%%%%%%%%%%%%%%%
Boolean Networks (BNs) are discrete dynamical systems $ X \equiv \{x_i\}$  of $N$ Boolean variables $x_i \in \{0,1\}$.
Interactions between variables are represented as a directed adjacency graph, the \emph{structural network}: $G=(X,E)$, where edges $e_{ji} \in E$ denote that variable $x_j$ is an input to variable $x_i$.
Furthermore, $X_i \equiv \{x_j \in X: e_{ji} \in E \}$ and $|X_i| = k_i$ denote the input set and the \emph{in-degree} of variable $x_i$, respectively.
Here, variables are updated synchronously according to deterministic logical functions:
$f_i: \{0,1\}^{k_i} \rightarrow \{0,1\}$, such that $x_i^{t+1} = f_i(X_i^t \subseteq X)$, where $X_i^t$ denotes the state of the inputs to $x_i$ at time $t\in\mathbb{N}$.

At time $t$, the network is in a \emph{configuration} of states $\bm{X}^t$, which is a vector of all variable states $x_i^t$ at $t$.
The set of all possible network configurations is denoted by $\mathcal{X} \equiv \{0,1\}^N$, where $|\mathcal{X}| = 2^N$.
The complete dynamical behavior of the system for all initial conditions is captured by the \emph{state-transition graph} (STG): $\mathcal{G}=(\mathcal{X}, \mathcal{T})$, where each node is a configuration $\mathbf{X}_{\alpha} \in \mathcal{X}$, and an edge $T_{\alpha,\beta} \in \mathcal{T}$ denotes that a system in configuration $\mathbf{X}_{\alpha}$ at time $t$ will be in configuration $\mathbf{X}_{\beta}$ at time $t+1$.
Under deterministic dynamics, only a single transition edge $T_{\alpha,\beta}$ is allowed out of every configuration node $\mathbf{X}_{\alpha}$.
Because $\mathcal{G}$ is finite, it contains at least one attractor, as some configuration or cycle of configurations must repeat in time \cite{Wuensche1998}.
An exemplar STG is shown in Figure 1 (top, left).

%%%%%%%%%%%%%%%%%%%%%%
% Figure with CSTG
\begin{figure}[t]
	\centering
	\includegraphics{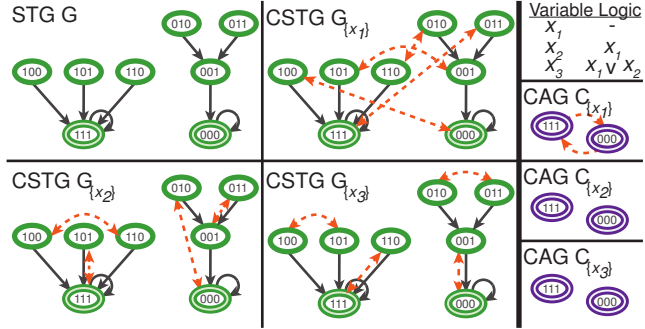}
	\caption[Controlled State Transition Graph]{The state transition graph (STG) and the controlled variants (CSTG) for an exemplar Boolean Network using the Feed-Forward network structure (Figure 2A), with the logical transition functions given in the upper right.  Configurations are shown as green nodes, attractors are highlighted green nodes, and transitions are illustrated as solid black arrows.  The CSTG $\mathcal{G}_{D}$ for the three singleton driver variable sets $D \equiv \{x_1\}, \{x_2\}, \{x_3\} $ are shown with controlled transitions denoted by dashed, orange arrows.  The controlled attractor graphs CAG $\mathcal{C}_{D}$ are also depicted for the singleton driver variable sets with the attractors shown as purple highlighted nodes and dashed orange arrows denoting the existence of at least one perturbed transition between attractor basins (if any exist).}
\end{figure}
%%%%%%%%%%%%%%%%%%%%%%%%

%%%%%%%%%%%%%%%%%%%%%%%%%%%
\subsection*{Control Measures}
%%%%%%%%%%%%%%%%%%%%%%%%%%%

We study the control exerted on the dynamics of a BN by a subset of \emph{driver variables} $D \subseteq X$.
Here, control \emph{interventions} are instantaneous bit-flip perturbations to the state of the variables in $D$ \cite{Willadsen2007}.
To capture all possible trajectories due to controlled interventions on $D$, we introduce the \emph{controlled state transition graph} (CSTG): $\mathcal{G}_D =(\mathcal{X}, \mathcal{T} \cup \mathcal{T}_D)$.
The CSTG is an extension of the STG, where a set of additional edges $\mathcal{T}_D$ denotes transitions from every configuration to each of its possible $2^{|D|}-1$ perturbed counterparts.
In Figure 1, three examples of CSTG are shown with interventions to only one of the three variables: $D = \{x_1\}, \{x_2\}, \{x_3\}$.

From the point of view of control theory \cite{Lin1974,Sontag1998}, the dynamics of a network of variables $X$ is \emph{controllable} by interventions to a subset of driver variables $D \subset X$ when every configuration is reachable from every other configuration in $\mathcal{G}_D$.
A configuration $\mathbf{X}_\beta$ is reachable from $\mathbf{X}_\alpha$ if a directed path from $\mathbf{X}_\alpha$ to $\mathbf{X}_\beta$ exists \cite{Sontag1998}.
For BN this is equivalent to requiring that the CSTG $\mathcal{G}_D$ be strongly connected.
To measure how much control $D$ can exert, we tally the fraction of configurations that are reachable by interventions to $D$.
Given a configuration $\mathbf{X}_\alpha$, the \emph{fraction of reachable configurations} $r(\mathcal{G}_D, \mathbf{X}_\alpha)$ is the number of other configurations $\mathbf{X}_\beta$ lying on all directed paths from $\mathbf{X}_\alpha$, normalized by the total number of other configurations $2^{N-1}$.
The \emph{mean fraction of reachable configurations}:
\begin{equation}
	\label{eq:frac_reached}
	\overline{R}_D = \frac{1}{2^{N}} \sum_{\mathbf{X}_\alpha \in \mathcal{X}} r(\mathcal{G}_D,\mathbf{X}_\alpha)
\end{equation}
measures the proportion of configurations which are on average reachable by controlling the set of driver variables $D$.
When a network is fully controlled by $D$, $\overline{R}_D = 1.0$, but for partially controlled networks $\overline{R}_D \in [0.0,1.0)$.

Notice that $\overline{R}_{\emptyset} \geq 0$, because the STG $\mathcal{G}$ of a network ($D \equiv \emptyset$) naturally contains transitions between configurations.
Therefore, it is useful to measure the control exerted by a set of driver variables $D$ beyond the uncontrolled dynamics.
To this end, we introduce the \emph{mean fraction of controlled configurations}:
\begin{equation}
	\label{eq:frac_controlled}
	\overline{C}_D =\overline{R}_D - \overline{R}_{\emptyset}
\end{equation}
It measures the fraction of configurations which are on average reachable by controlling the driver variables $D$ that were not already reachable via the natural dynamics.  By definition, $\overline{C}_D\leq\overline{R}_D$ for any system and set of driver variables.

In practice, only certain subsets of configurations are meaningful.  These subsets are typically cast as either attractors for the system dynamics or specific trajectories through the state space.
Consider the case of BNs as models of biochemical regulation; attractors represent different cell types \cite{Kauffman1993,Muller2011,Kauffman1969}, diseased or normal conditions \cite{Zhang2008}, and wild-type or mutant phenotypes \cite{MarquesPita2013}.
In this context, the formal sense of controllability is well beyond what is necessary.
What is most relevant for some systems is to uncover the driver variables which can steer dynamics from attractor to attractor; transient configurations are irrelevant.

To measure this more realistic sense of control, we introduce the \emph{controlled attractor graph} (CAG): $\mathcal{C}_D = (\mathcal{A},\mathcal{B}_D)$.
In this graph, each node $\mathbf{A}_{\kappa} \in \mathcal{A}$ represents an attractor.
A \emph{basin edge} $b_{\kappa\gamma} \in \mathcal{B}_D$, denotes the existence of at least one path from attractor $\mathbf{A}_{\kappa}$ to attractor $\mathbf{A}_{\gamma}$.
In Figure 1 (right-side), three examples of CAGs are shown.
The \emph{mean fraction of reachable attractors} is then given by
\begin{equation}
\label{eq:frac_attract_reach}
	\overline{A}_D =\frac{1}{|\mathcal{A}|} \sum_{\mathbf{A}_\kappa \in \mathcal{A}} r(\mathcal{C}_D,\mathbf{A}_{\kappa})
\end{equation}
where $\kappa=1 \ldots |\mathcal{A}|$.
It measures the fraction of attractors which are on average reachable by controlling the driver variables in $D$.  A network which can be controlled from any of its attractors to any of its attractors must have $\overline{A}_D = 1.0$; when $D \equiv \emptyset$, all attractors reside in disconnected basins in the original STG so $\overline{A}_{\emptyset} = 0.0$.
Naturally, if a network is fully controllable by $D$ in the control theory sense ($\overline{R}_D = 1.0$), $\overline{A}_D = 1.0$.

%%%%%%%%%%%%%%%%%%%%%%%%%%%
\section*{Control Portraits of Complex Systems}
%%%%%%%%%%%%%%%%%%%%%%%%%%%

%%%%%%%%%%%%%%%%%%%%%%%%%%%
\subsection*{Boolean Network Ensembles}
%%%%%%%%%%%%%%%%%%%%%%%%%%%

Given the structural network $G=(X,E)$ for a BN, many different logical functions $f_i$ can be assigned to each Boolean variable $x_i$ (see Background).
An \emph{ensemble} of BNs is constructed by considering all possible logical functions constrained by the fixed structure $G$ \cite{Kauffman2003,Ciliberti2007} (see SM).
However, since non-contingent functions (e.g.\ tautology and contradiction) are not found in most biological models, we divide the full ensemble into \emph{contingent} and \emph{non-contingent} subsets as follows: those BNs which only contain contingent functions and those BNs which contain at least one non-contingent transition function (NC).

Within the set of contingent functions, there are \emph{canalizing} functions which depend only on a subset of their input variables \cite{Kauffman2004,Reichhardt2007}.
These functions are ubiquitous in BN models of gene regulation and contribute mechanisms of functional redundancy and degeneracy \cite{MarquesPita2013,Gershenson2006}.
The redundancy of some logical functions means that the \emph{effective} structure of interactions is reduced \cite{MarquesPita2013,MarquesPita2015}: some edges of the structural graph $G$ play no role in determining the transitions between configurations.

Since control methodologies based on network structure assume that all interactions (edges) in the structural network are relevant for system dynamics, we further subdivide the contingent subset into two disjoint subsets:
BNs which contain fully canalizing functions and thus possess a \emph{reduced effective structure} (RES), and those without canalizing functions retaining a \emph{full effective structure} (FES).
Naturally, the FES subset is the scenario most coherent with the idea of using structure to predict controllability, since all interactions in the underlying structural graph $G$ are dynamically relevant.

%%%%%%%%%%%%%%%%%%%%%%
% Figure with motif structure
\begin{figure}[t]
	\centering
	\includegraphics{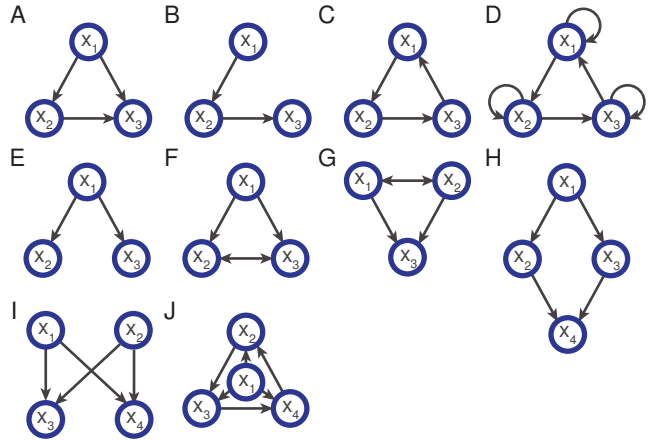}
	\caption[Exemplar Network Motifs]{Directed network structure motifs used in this study: A) Feed-Forward motif, B) Chain motif, C) Loop motif,  D) Loop motif with self-interactions, E) Fan motif, F) Co-regulated motif, G) Co-regulating motif, H) BiParallel motif, I) BiFan motif, and J) Dominated Loop motif. }
\end{figure}
%%%%%%%%%%%%%%%%%%%%%%

%%%%%%%%%%%%%%%%%%%%%%%%%%%
\subsection*{Network Motifs}
%%%%%%%%%%%%%%%%%%%%%%%%%%%

We first consider the entire ensemble of BNs with simple structural graphs known as \emph{network motifs} \cite{ShenOrr2002}.
These prototype networks have been useful for exploring the relationship between structure and dynamics of complex networks \cite{Ingram2006, Prill2005}.
The motifs considered in our analysis are depicted in Figure 2.

Consider the Feed-Forward network motif of $N=3$ variables \cite{Mangan2003} shown in Figure 2A.
In this case, the full ensemble consists of $64$ distinct BNs of which $36$ are NC, $8$ have RES, and $20$ have FES.
Figure 1 depicts the logic of one FES network instance for this motif, along with its STG, CSTGs, and CAGs for various driver sets $D$.
The control portrait of the full BN ensemble is shown in Figure 3; control measures $\overline{R}_{D}$ and $\overline{C}_{D}$ are shown for all possible driver sets of one or two variables.

Using solely this motif's interaction network, structural control (both the SC and MDS methods) predicts that variable $x_1$ is capable of fully controlling the network.
However, our analysis reveals that this driver variable can fully control only $8$ networks from the ensemble ($4$ RES and $4$ FES), while the other $56$ BNs (with the same structure) are not fully controlled (Figure 3).
It is noteworthy that even when considering the FES subset --- the scenario most coherent with the idea of using structure to predict the controllability of the dynamics --- only $4$ out of $20$ BNs are fully controlled by interventions on $x_1$.
It is clear that even in the case of such a simple motif, structure does not predict the control of dynamics.
An extended analysis of the controlled Feed-Forward BN ensemble is provided in the SM.

%%%%%%%%%%%%%%%%%%%%%%
% Figure of FeedForward Control Analysis
\begin{figure}[t]
	\centering \includegraphics{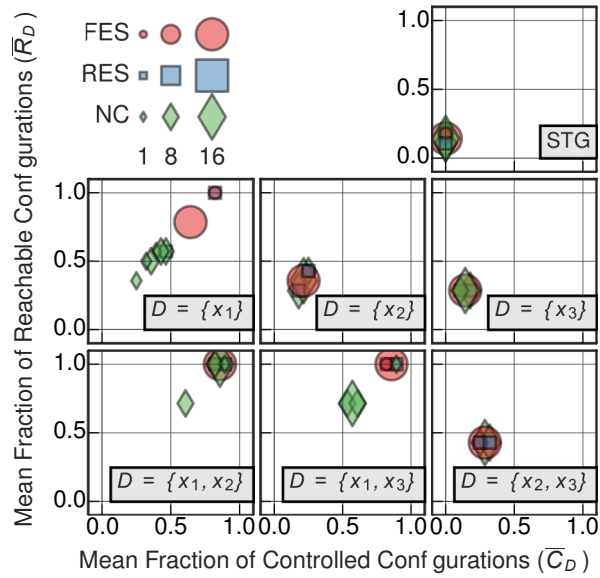}
	\caption[Control Portrait for the Feed-Forward Motif]{Control portrait of the BN ensemble constrained by the Feed-Forward network motif.  The mean fraction of reachable configurations $\overline{R}_D$ and the mean fraction of controllable configurations $\overline{C}_D$ for the full ensemble of $64$ BNs with structure given by the Feed-Forward network motif shown in Figure 2A, as controlled by all driver variable sets of one or two variables.  The full effective structure (FES) subset is highlighted by red circles, the reduced effective structure (RES) subset is shown in blue squares, and the non-contingent subset (NC) is shown by green diamonds; the area of the object corresponds to the number of networks at that point.}
\end{figure}

%%%%%%%%%%%%%%%%%%%%%%

Let us now consider the $N=3$ variable loop motif with self-interactions (Figure 2D).  The full ensemble of BNs constrained by this motif is much larger than the previous example (every variable has $k_i=2$ inputs); it consists of $4096$ networks of which $1352$ are NC, $1744$ have RES, and $1000$ have FES.
Figure 4A shows the control portrait of this motif's BN ensemble for a single $(D\equiv\{x_i\})$ or pair $(D\equiv\{x_i , x_j\})$ of driver variables.
The control portrait of the STG illustrates the difference between the two measures of controllability. While $\overline{R}_D$ varies greatly, $\overline{C}_D = 0$ for all BNs. This means that in some BNs, many configurations can be reached simply because the transient dynamics move through many network configurations.
Structural control methodologies ignore this natural propensity for control (self-organization).
Thus we use the measure $\overline{C}_D$ to tally only the proportion of transitions that result from control interventions.

The control portraits in Figure 4 again demonstrate that structure fails to characterize network control.
In this case, SC predicts that any single variable is sufficient for full controllability, while MDS requires any two variables to achieve the same.
Yet controllability varies greatly for both cases, depending on the particular transition functions of each BN in the ensemble.
For $77\%$ of the BNs in the ensemble a single variable is not capable of fully controlling dynamics; even two-variable driver sets fail to control $44\%$ of the BNs.

Similar results hold for the mean fraction of reachable attractors ($\overline{A}_D$) shown in Figure 4B (middle, right).
For $36\%$ of the BNs in the ensemble, a single variable is not capable of fully controlling the system between attractors; even two-variable driver sets fail to control $20\%$ of the BNs, regardless of the dynamical subset.
Discounting the $1868$ networks (Figure 4B, left) with only one attractor (hence $\overline{A}_D=1$) further emphasizes the variation in attractor control, increasing the above proportions to $65\%$ and $36\%$ for one and two driver variables, respectively.
Therefore, even if we analyze controllability from the point of view of attractor control rather than the stringent criteria of full controllability, single- and two-variable driver sets fail to achieve controllability of all networks in this ensemble.

The control portraits of the other network motifs analyzed are presented in the Supplemental Material.
Their analysis supports the same conclusion: predictions made from structure-only methods are only true for a small number of possible BNs.  In general, they fail to predict the actual controllability of all the BN dynamics that can occur for a given motif structure.

%%%%%%%%%%%%%%%%%%%%%%
% Figure of Loop Motif with self-interactions Analysis
\begin{figure}[ht]
	\centering
	\includegraphics{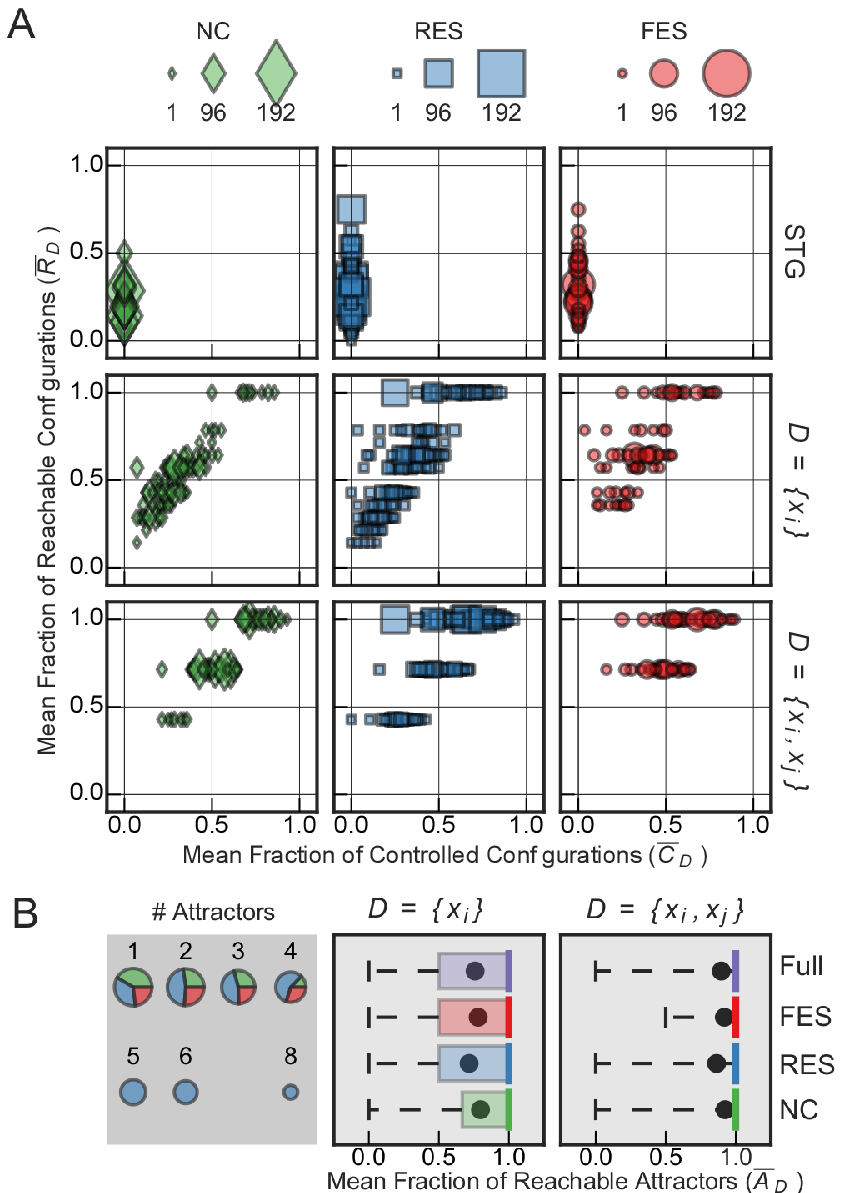}
	\caption[Control Portrait for the Loop Motif with self-interactions]{Control portrait of the BN ensemble constrained by the Loop network motif with self-interactions. A) The mean fraction of reachable configurations $\overline{R}_D$ and the mean fraction of controllable configurations $\overline{C}_D$ for the full ensemble of $4096$ BNs with structure given by the Loop network motif with self-interactions shown in Figure 1D, as controlled by the driver variable sets $D\equiv\emptyset$ (STG), $D\equiv\{x_i\}$, and $D\equiv\{x_i,x_j\}$ (due to the symmetry of the network, all sets of size one are equivalent, likewise those of size two).  The full effective structure (FES) subset is shown by red circles, the reduced effective structure (RES) subset is shown in blue squares, and the non-contingent (NC) subset is shown by green diamonds; the area of the object corresponds to the number of networks at that point.  B) (left) The number of attractors for each network in the full ensemble spans from $1-8$, the area of each pie chart scales logarithmically with the number of attractors, from $1868$ to $1$; the colored slices delineate the subset decompositions for NC, RES, and FES.  (middle and right) Box plots for the distribution of the mean fraction of reachable attractors $\overline{A}_D$ for $D\equiv\{x_i\},\{x_i,x_j\}$ for the full ensemble (purple), NC, RES, and FES subsets.  In each case, the box shows the interquartile range, the median is given by the solid vertical line, the mean is given by the black circle, and the whiskers show the support of the distribution.}
	\end{figure}
%
%%%%%%%%%%%%%%%%%%%%%%

%%%%%%%%%%%%%%%%%%%%%%%%%%%
\subsection*{Models of biochemical regulation}
%%%%%%%%%%%%%%%%%%%%%%%%%%%

To better understand the interplay between structure and dynamics in the context of controlling complex systems, we study three BN models from systems biology which are considerably larger than the network motifs of the previous section.

\subsubsection*{Drosophila melanogaster}
During the early ontogenesis of the fruit fly, the specification of adult cell types is controlled by a hierarchy of a few genes.
The Albert and Othmer segment polarity network (SPN) is a BN model \cite{Albert2003} capable of predicting the steady-state patterns experimentally observed in wild-type and mutant embryonic development with significant accuracy.
Here, we analyze the single-cell SPN consisting of $17$ gene and protein variables (see SM).

%%%%%%%%%%%%%%%%%%%%%%
% Figure of Drosophila control analysis
\begin{figure}[ht]
	\centering
	\includegraphics{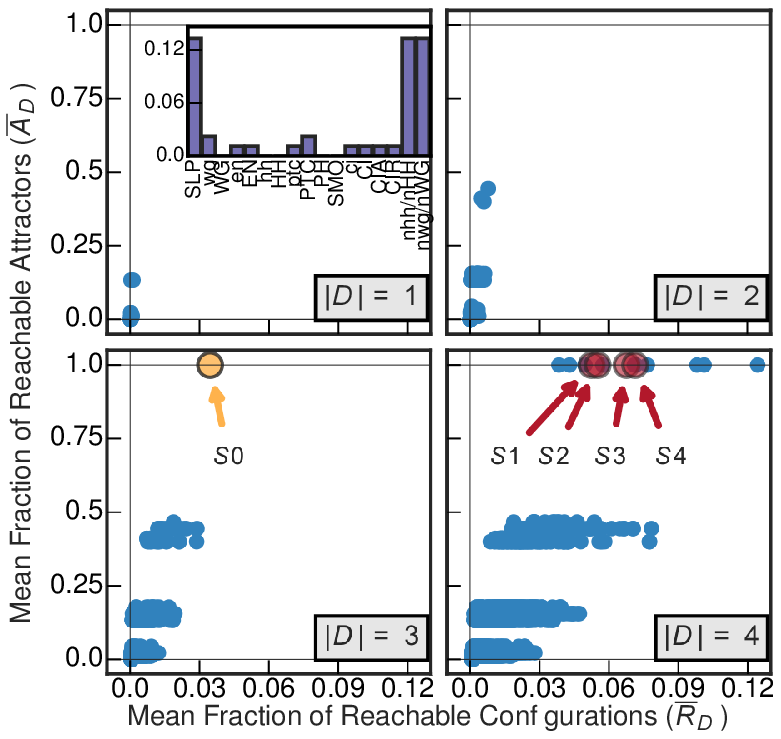}
	\caption[Control Analysis for \emph{Drosophila melanogaster}]{Control of the single-cell segment polarity network (SPN) of gene and protein regulation in \emph{Drosophila melanogaster} for all driver variable subsets of size $|D| = 1$, $|D| = 2$, $|D| = 3$, and $|D| = 4$.  (inset) The mean fraction of reachable attractors $\overline{A}_D$ for each singleton driver variable set. The driver subsets predicted by structural controllability (SC) to fully control the network are highlighted in red and labeled $\mathcal{S}1$, $\mathcal{S}2$, $\mathcal{S}3$, $\mathcal{S}4$.  The three variable driver subset with full attractor control is highlighted in yellow and labeled $\mathcal{S}0$ (see SM for further details).}
\end{figure}
%%%%%%%%%%%%%%%%%%%%%%

Previous analysis has shown that the SPN model is controlled by the upstream value of the Sloppy Pair Protein (SLP) and the extra-cellular signals of the Hedgehog and Wingless proteins from neighboring cells nhh/nHH and nWG \cite{Albert2003}.
The control portrait of this model also demonstrates that these three variables (driver set $\mathcal{S}0$ in Figure 5) are capable of fully controlling the dynamics from any attractor to any other attractor.
This is to be expected in segment polarity regulation since it is a highly orchestrated developmental process.
The attractor control ability of individual nodes of the SPN in the inset of Figure 5 further highlights this behavior, only the $3$ chemical species mentioned above have a high $\overline{A}_D$ when controlled alone, while all internal variables have negligible influence.

The SC analysis of the SPN's structural graph identifies $4$ subsets of $|D| = 4$ driver variables, indicated in Figure 5 by enlarged red circles and labeled $\mathcal{S}1, \mathcal{S}2, \mathcal{S}3$ and $\mathcal{S}4$ (details in SM).
$\mathcal{S}0$ is a subset of these $4$ variable subsets, so naturally they also achieve $\overline{A}_D = 1$, but they all include an additional variable which is redundant for this purpose.
However, none of these subsets are sufficient for fully controlling the BN as predicted by SC,
these driver sets can control dynamics only to a very small proportion of configurations; $\overline{R}_{D \equiv \mathcal{S}4}\approx 0.071$ is the maximum value attained.
These $4$ driver sets also show considerable variation in $\overline{R}_D$, demonstrating that predictions with equivalent support from the point of view of the SC theory, lead to distinct amounts of real controllability.
Interestingly, there are $5$ driver variable sets of size $|D| = 4$ that lead to greater controllability (with a maximum of $\overline{R}_D \approx \overline{C}_D \approx 0.124$) than predicted by SC.  Thus, SC fails to even correctly predict the $4$-variable driver sets with greatest controllability.

The MDS analysis of the SPN model predicts that $|D| = 7$ variables are required to fully control the system dynamics and uncovers $8$ equivalent driver variable sets of this size (see SM).
Not surprisingly, all of the MDS driver variable sets achieve full attractor control ($\overline{A}_D=1$) since they contain $\mathcal{S}0$; however, none can fully control the network dynamics achieving only a maximum $\overline{R}_D\approx0.31$.
Thus, the driver sets predicted by both SC and MDS are not sufficient to control dynamics in the control theory sense, and predict more variables than necessary to achieve attractor control.

%%%%%%%%%%%%%%%%%%%%%%
% Figure of Yeast control analysis
\begin{figure}[ht]
	\centering
	\includegraphics{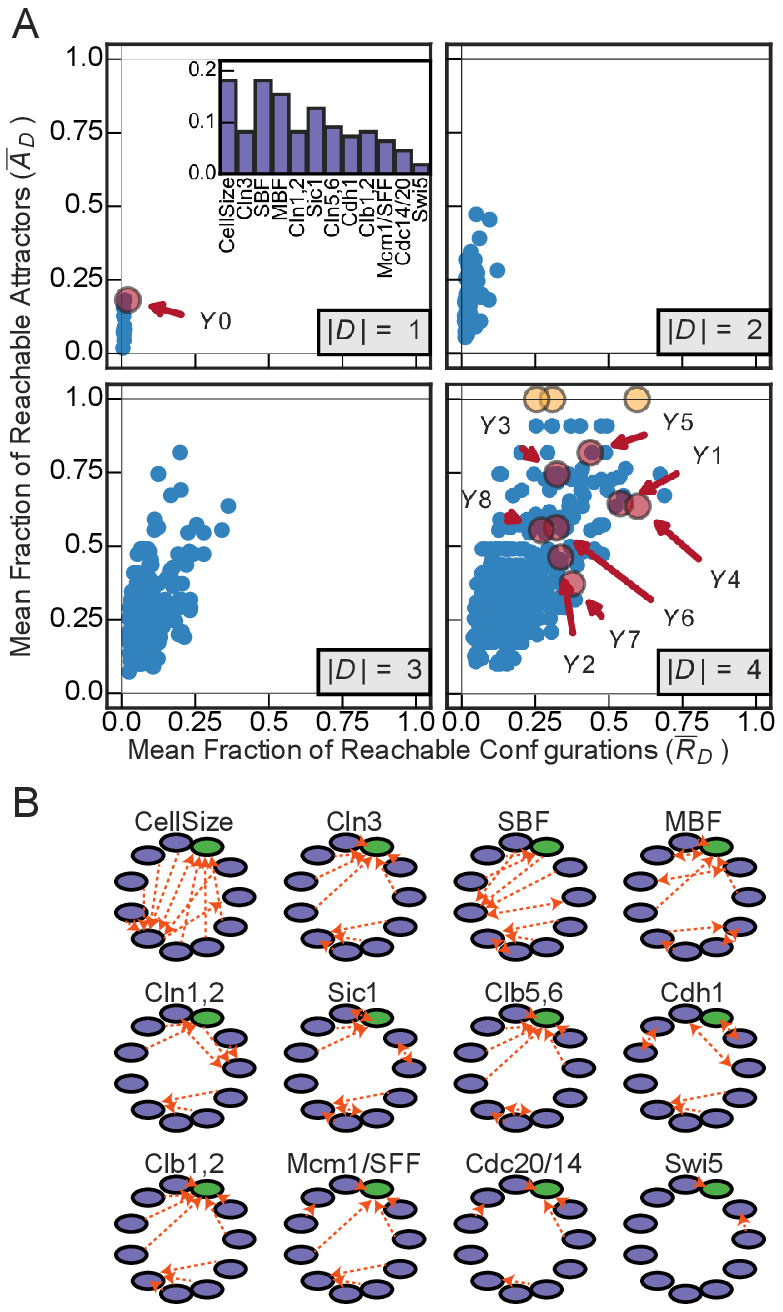}
	\caption[Control Analysis for \emph{Saccharomyces cerevisiae}]{A) Control of the eukaryotic cell cycle of budding yeast \emph{Saccharomyces cerevisiae} (CCN) for all driver variable subsets of size $|D| = 1$, $|D| = 2$, $|D| = 3$, and $|D| = 4$.  (inset) The mean fraction of reachable attractors $\overline{A}_D$ for each singleton driver variable set.  The subset predicted to fully control the network are highlighted in red and labeled $\mathcal{Y}0$ for structural controllability (SC), while those predicted by minimum dominating sets (MDS) are labeled $\mathcal{Y}1-8$.  The driver variable subsets with full attractor control are highlighted in yellow (see SM for further details).  B) Controlled Attractor Graphs (CAGs) for each singleton driver variable set.  The wild-type attractor is highlighted in green, all other attractors are in purple.}
	\label{fig:yeast}
\end{figure}
%%%%%%%%%%%%%%%%%%%%%%

\subsubsection*{Saccharomyces cerevisiae}
The eukaryotic cell cycle process of the budding yeast \emph{Saccharomyces cerevisiae} reflects the cyclical gene expression activity that leads to cell division.
Here, we use the $12$ variable simplified Boolean model of the yeast Cell-Cycle Network (CCN) derived by Li et al. \cite{Li2004}.
The SC analysis of the CCN interaction graph identifies only one driver variable ($\mathcal{Y}0=\{CellSize\}$) to be sufficient for fully controlling the BN's dynamics.
Yet, as demonstrated in Figure 6A, it only achieves negligible configuration control ($\overline{R}_{\mathcal{Y}0}\approx 0.021$) and very weak attractor control ($\overline{A}_{\mathcal{Y}0}\approx 0.19$).
Similarly, MDS analysis identifies $8$ driver variable sets of size $|D|=4$ ($\mathcal{Y}1$ to $\mathcal{Y}8$), none of which achieve full control.
It is particularly interesting that the driver sets predicted by MDS lead to values of both $\overline{A}_D$ and $\overline{R}_D$ that are essentially random, demonstrating once again that predictions with equivalent support from the point of view of the structure-only theories lead to widely different amounts of real controllability. 
Our analysis finds $3$ driver sets of $|D|=4$ variables that achieve full attractor control (highlighted in yellow in Figure 6A and detailed in SM). 
Neither SC nor MDS predict those specific driver sets, which ultimately provide the most useful form of control in such systems.
Unlike the SPN, there are no ``chief controller'' variables in this network, as most variables achieve a similar value of $\overline{A}_D$ when controlled alone (see inset in Figure 6A).

The CCN was designed such that there is a large attractor basin towards a wild-type attractor which is robust to perturbation \cite{Li2004, Willadsen2007}.
However, our analysis illuminates the tradeoff between robustness and flexibility in relation to system controllability.
While a large basin of attraction facilitates controlling the system towards the wild-type behavior (high wild-type robustness), it also reduces the ability to control the system to other smaller basins of attraction (mutant phenotypic behavior), reflecting a tradeoff between wild-type robustness and low flexibility for potential evolvability (a property that was not initially designed into the model to begin with).
This tradeoff is further elaborated by the CAGs for all single-variable driver sets, shown in Figure 6B. Some variables have a propensity to control the system towards the wild-type attractor (green node) or allow the system to remain there (e.g. $Cln3$, $Clb5,6$, $Clb1,2$, $Mcm1/SFF$, $Cdc20/14$), while only a few can control the system out of this attractor (e.g. $CellSize$, $SBF$, $Cln1,2$).  See SM for more details.

A third model of biochemical regulation in the floral organ arrangement in the flowering plant \emph{Arabidopsis thaliana} was analyzed, leading to a similar failure to predict actual control (see SM for details).

\subsection*{Canalization and Controllability}

When fully canalizing functions are present in a BN, not all of the edges in the structural graph contribute to the collective dynamics; there exists a subgraph that fully captures the dynamically relevant interactions (an \emph{effective} structural graph) \cite{MarquesPita2013,MarquesPita2015}.
Moreover, most Boolean functions are \emph{partially} canalizing \cite{MarquesPita2013,Reichhardt2007} whereby in some input conditions a subset of inputs is redundant, but in other conditions it is not.
This means that most edges in the underlying structural graph of a random BN are either entirely or partially redundant.

Since structural controllability methods assume that every edge of the underlying structure fully contributes to the dynamics, it is reasonable to suspect that the larger the \emph{mismatch} between the structural graph and the effective structural graph, the more the predictions from SC and MDS will fail.
To study this hypothesis, we constructed several ensembles of BNs where there is a perfect match between the structural graph and the effective structure graph.

First consider the ensemble of BNs with the structural graph of the CCN, but with transition functions chosen from the set of two non-canalizing functions that exist for each variable's in-degree. 
This constitutes a \emph{Full Effective Connectivity} (FEC) ensemble of BNs whose effective structure perfectly matches the original structural graph of the CCN --- there is no canalization in the dynamics of these networks.

Even though both SC and MDS fail to predict controllability correctly for a sample of $50$ networks from the FEC ensemble, our analysis reveals that they are more easily controlled by smaller driver sets than the original CCN model.
Specifically, $\overline{R}_D$ and $\overline{A}_D$ averaged over all driver variable sets is larger for every FEC sample than for the original CCN model (details in SM).
Many networks in the FEC ensemble were fully controllable by $2$ driver variables and all networks could be fully controlled by $3$ driver variables---whereas the original CCN requires $4$ variables for full attractor control.

Interestingly, canalization can also be used to improve controllability if selected appropriately.  To see how, we compare BN ensembles with no canalization whatsoever to those with only fully canalizing functions for each motif (see SM for details).
This uncovers the cases where canalization actually improves BN controllability, even beyond the controllability attained by networks with no canalization.
In all such cases, the resulting effective structure reduces the original structural graph to simpler linear chain motifs (Figure S36 in SM). This way, canalization of the individual variable transition functions is orchestrated to obtain pathways that channel the collective dynamics towards greater control (macro-level canalization \cite{MarquesPita2013}).
Because these linear chain effective structures match the assumptions of structure-only methods more accurately, their predictions are correct in such cases.
Thus, canalization can enhance the accuracy of structure-only control methodologies if transition functions are appropriately selected to reduce the effective structure to a linear chain.
Naturally, when the size of the network increases from simple motifs to realistic networks, BNs with such precise effective structure become extremely rare in the ensembles.

%%%%%%%%%%%%%%%%%%%%%%%%%%%
\section*{Discussion}
%%%%%%%%%%%%%%%%%%%%%%%%%%%

We studied the interplay between structure and dynamics in the control of complex systems using ensembles of BNs and existing models of biochemical regulation.
The analysis of the BN ensembles constrained by network motifs demonstrates that structure-only methods fail to properly characterize control; there is a large variation of possible dynamics that can occur for even the simplest network.
The situation only gets worse for structure-only methods when we scale up to real models of biochemical regulation. Our analysis demonstrates that structural control predictions can both underestimate or overestimate the number of driver variables in these systems.
These approaches also fail to predict which sets of variables best control dynamics as evaluated by: how much of the total configuration space is accessible ($\overline{R}_D$), how much of the configuration space is accessible beyond the natural system dynamics ($\overline{C}_D$), and the ability to transition between attractors ($\overline{A}_D$).
Often, arguments made about how easy it is to control network types (e.g. biological vs. social \cite{Egerstedt2011}) hinge on how many driver variables are predicted by structural control theories.
Yet, our analysis reveals that much variation in real control occurs for the same structure and number of driver variables.

Our approach also lays the groundwork for understanding which restrictions must be enforced on the transition functions of BNs such that structure may suffice for predicting controllability or at least improve the accuracy of structure-only methods in predicting control.
In our experiments with ensembles of network motifs, canalizing transition functions generally rendered structure-only methods less effective at predicting the control of dynamics. Given the generality of motifs as network building blocks, this suggests our results will generalize to larger systems, as already observed in the three larger gene-regulation models considered here.
On the other hand, we showed that it is possible to orchestrate canalization such that the effective structure matches the assumptions of structure-only methods, leading to more accurate predictions about control.
This effect was identifiable in small networks, where it is easy to find the necessary effective structures, however, such structures are rare in the space of all possible dynamics for larger networks.
Nonetheless, in principle, evolution or human design could select for such networks. 

Crucially, without more information about variable dynamics, we certainly cannot assume that a given multi-variate dynamical system meets the assumptions of structure-only methods.
For instance, the CCN model uses canalization to make controllability harder than predicted by structure-only methods, while the SPN model uses canalization to control dynamics to the wild-type attractor more easily than suggested by the same methods.
All this suggests that canalization plays an important, nontrivial role in determining structure-dynamic relationships.  Further research can explore this interplay in greater detail. But our current analysis suggests that, without more information about variable dynamics, structure-only methods cannot be accepted as even an approximation of how control occurs in complex systems.

The control measures we introduced here for BNs provide a complementary viewpoint to those developed to study system robustness \cite{Chaves2006, Willadsen2007}. 
Both concepts are based on the response of the system to perturbations.
However, robustness focuses on the quantity of perturbations to which the system's dynamics is invariant, whereas control tracks the perturbations which alter the system's dynamics.
Future research will also explore other characteristics of the controlled state transition graph and controlled attractor graph so that the relationship between robustness and control can be better studied.

Boolean Networks are ideal, parsimonious systems for our study since they are defined by both a clear interaction structure and rich nonlinear dynamics using only binary variables.
However, our conclusions are not necessarily limited to this type of network.
The control measures used in our study are formulated with respect to a state transition graph, and are therefore applicable to any discrete, deterministic dynamical system. Our conclusions are thus likely to extend to other classes of complex systems.
Indeed, several recent papers have also questioned the validity of structure-only arguments for control of other non-linear systems \cite{Motter2015}.  These arguments are grounded in the treatment of finite time constants and self-interactions \cite{Cowan2012}, the numerical limitations of nonlocal controlled trajectories \cite{Sun2013}, or the role of symmetry in the non-linear dynamics \cite{Whalen2015}.
Understanding the discrepancy between network structure and control is also important for specific applications where methods which construct a specific controller (i.e. an algorithm that identifies a specific sequence of controlled interventions given a set of constraints) are desired.  
Structure-only predictions do not aim to predict controllers, rather they focus on the mere identification of driver variables.  
The identification of controllers is the subject of much research in systems biology and complex systems; in this case, a greater disparity between structure-only predictions and actual control is expected \cite{Motter2015}.

Ultimately, methodologies that can help us predict control in complex networks while avoiding computational complexity should be developed, but they must combine characteristics of both the structural and dynamical properties of the system.
Promising methods are already being developed which include both structure and dynamics, such as monotone control systems \cite{Angeli2003}, master stability functions \cite{Gutierrez2012}, schema redescription \cite{MarquesPita2013}, and stabilization subgraphs \cite{Zanudo2015}.
Understanding how such simplifications scale-up while providing a reasonable account of how control operates is very important, especially in real-world systems.
This can be accomplished via the type of study we undertook here to analyze the effectiveness of structure-only methods in predicting the controllability of complex systems.

%%%%%%%%%%%%%%%%%%%%%%%%%%%
\bibliography{GatesRochaControlComplexSystemsRefs}

%%%%%%%%%%%%%%%%%%%%%%%%%%%
\section*{Acknowledgements}

We thank Randall Beer, Artemy Kolchinsky, Santosh Manicka, Eran Agmon, Ian Wood, and three anonymous reviewers for helpful conversations and feedback. This work was partially supported by a grant from the National Institutes of Health, National Library of Medicine Program, grant 01LM011945-01 "BLR: Evidence-based Drug-Interaction Discovery: In-Vivo, In-Vitro and Clinical," a fellowship from the NSF IGERT "The Dynamics of Brain-Body-Enivronment Systems in Behavior and Cognition", a grant from the Fundacao para a Ciencia e a Tecnologia (Portugal), PTDC/EIA-CCO/114108/2009 ?Collective Computation and Control in Complex Biochemical Systems?, as well as a grant from the joint program between the Fundacao Luso-Americana para o Desenvolvimento (Portugal) and the National Science Foundation (USA), 2012-2014, "Network Mining For Gene Regulation And Biochemical Signaling."  The funders had no role in study design, data collection and analysis, decision to publish, or preparation of the manuscript.  The source code is available upon request.

\section*{Author contributions statement}
A.G. and L.R. designed the study, analyzed the results, and wrote the manuscript. 

\section*{Additional information}
The authors declare no competing financial interests.

\includepdf[pages=-, offset=15 -15]{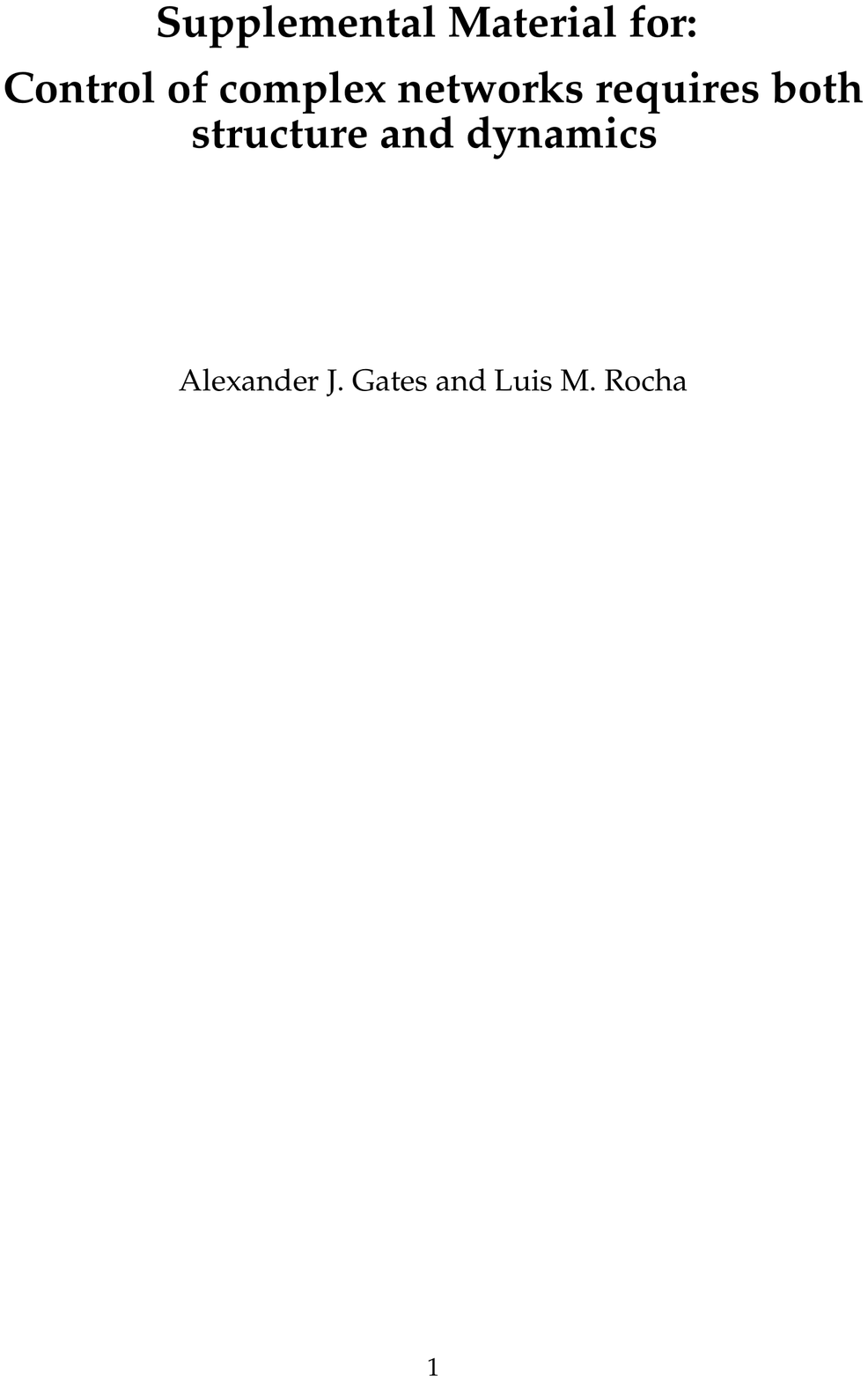}

\end{document}